\documentstyle[12pt]{article}
\begin{document}
\begin{center}
{\bf About Some Distinguishing Features of the Weak Interaction}\\
\vspace{2cm}
Kh. M. Beshtoev \\
\vspace{2cm}
Joint Institute for Nuclear Research, Joliot Curie 6\\
141980 Dubna, Moscow region, Russia
\vspace{2cm}
\end{center}
\par
{\bf Abstract} \\

In this work it is shown that,
in contrast to the strong and electromagnetic theories,
additive conserved numbers (such as lepton, aromatic and another numbers)
 and $\gamma_5$ anomaly do not appear in the standard weak interaction theory.
It means that in this interaction the additive numbers cannot be conserved.
These results are the consequence of
specific character of the weak interaction: the right components of spinors
 do not participate in this interaction.
The schemes of violation of the aromatic and lepton numbers were considered.
\par
\noindent
Ketwords: weak interaction, additive numbers, $\gamma_5$-anomaly
\par
{\bf 1. Introduction}\\

Strong and electromagnetic interactions theories are left-right systemic
theories (i.e. all components of the spinors participate in these interactions
symmetrically). In contrast to this
only the left components of fermions participate
in the weak interaction. This work
is dedicated to  some consequences deduced from this specific
feature of the weak interaction.\\

\par
{\bf 2. Distinguishing Features of Weak Interactions}\\

\par
As it is well known from the Neuter theorem [1],
conserving currents appear
at global and local abelian and nonabelian gauge transformations.\\
\par
These values for local gauge transformations are: \\
electromagnetic current--$j^\mu$
$$
j^\mu = e \bar \Psi \gamma^\mu \Psi  ,
\eqno(1)
$$
where $e$ is an electrical charge;\\
the current of strong interactions--$j^{a \mu}$
$$
j^{a \mu} = q \bar \Psi T^a \gamma^\mu \Psi  ,
\eqno(2)
$$
where $q$ is charge of a strong interactions, $T^a$ is $SU(3)$ matrix, $a$ is
color.
\par
The currents $S^\mu_i$ obtained from  the global abelian transformation is
$$
S^\mu_i = i (\bar \Psi_i \partial_\mu \Psi_i) ,
\eqno(3)
$$
(where $i$ characterizes the type of the gauge  transformation) and
the corresponding
conserving current (the forth component of $S^\mu_i$) is
$$
I_i =  \int S^0_i d^3x = \int \epsilon \bar \Psi_i \Psi_i d^3x  ,
\eqno(4)
$$
where $\epsilon$ is the energy of fermion $\Psi_i$.\\
\par
Then conserving values of global gauge transformations are: \\
electrical number $Q$
$$
Q = \int e \epsilon \bar \Psi \Psi d^3 x  ;
\eqno(5)
$$
baryon numbers $B$
$$
B = \int \epsilon \bar \Psi_B \Psi_B d^3 x  ;
\eqno(6)
$$
the lepton numbers $l_i (i = e,\mu,\tau)$
$$
l_i = \int \epsilon \bar \Psi_{l_i} \Psi_{l_i} d^3 x  ;
\eqno(7)
$$
aromatic numbers and etc.

\par
In the vector (electromagnetic and strong interactions) theories all
components of spinors ($\Psi_L, \Psi_R$) participate in interactions.
In contrast to the strong and electromagnetic theories,
the right components of the spinors ($\bar \Psi_R, \Psi_R$)
do not participate in the weak interaction, i. e. this interaction does not
refer to the
chiral theory (in the chiral theory the left and right components of fermions
participate in the interaction in the independent manner).
Such character of the weak interaction leads to certain consequences:
impossibility  to generate
fermion masses [2] and to the problem of jointing this interaction to
the strong and electromagnetic interactions [3].
\par
Let us consider another consequences of this specific
feature of the weak interaction.
\par
The local conserving current $j^{\mu i}$
of the weak interaction has the following form:
$$
j^{\mu i} = \bar \Psi_L \tau^i \gamma^\mu \Psi_L  ,
\eqno(8)
$$
where $\bar \Psi_L, \Psi_L$ are lepton or quark doublets
$$
\left(\begin{array}{c} e\\ \nu_e \end{array}\right)_{i L}
$$
$$
\left(\begin{array}{c} q_1\\ q_2 \end{array}\right)_{i L},\qquad i = 1-3  .
$$
\par
If now we take into account that in the right components
of fermions $\bar \Psi_{i R}, \Psi_{i R}$ do not participate in the
weak interaction, then from (4) for abelian currents we get
$$
I_i = \int \epsilon \bar \Psi_{i L} \Psi_{i L} d^3 x  \equiv 0  ,
\eqno(9)
$$
i.e. (in contrast to the strong and electromagnetic
interactions) no conserving additive numbers appear in the
weak interaction.
\par
It is clear that the lepton and aromatic numbers appear outside
the weak interaction and it is  obvious that the interaction, where these
numbers appear, must be a left-right symmetric one.
\par
It is also clear that, since in the weak interaction no
conserving additive numbers appear, then the additive
(aromatic, lepton and etc.) numbers can be violated
in the weak interaction . Thus, the violation scheme of aromatic numbers,
as well known, is Cabibbo-Kobayashi-Maskawa
matrices [4, 5]
$$
V =
\left( \begin{array}{ccc} V_{ud}& V_{us}& V_{ub}\\ V_{cd}& V_{cs}& V_{cb}\\
V_{td}& V_{ts}& V_{tb} \end{array} \right)  ,
\eqno(10)
$$
where $u, d, s, c, b, t$ are quarks.
\par
The analogous scheme can be used to discribe of the violation of the
lepton numbers [5]
$$
V =
\left( \begin{array}{ccc} X_{ee}& X_{e\mu}& X_{e\tau}\\ X_{\mu e}&
X_{\mu\mu}& X_{\mu\tau}\\ X_{\tau e}& X_{\tau\mu}& X_{\tau\tau} \end{array}
\right) ,
\eqno(11)
$$
where $e, \mu, \tau$ are leptons.
\par
It is necessary to stress that, probably, in the weak interaction there is no
conserving baryon number $B$,
third projection the weak isospin ($I_3^w$),
and etc. ( see Eqs (7),(9)),
but it leads to any consequences since
the local electric, strong and weak currents are conserved.
\par
All the above considered violations of these numbers in the weak interaction
are the direct violations.
\par
Now we consider the problem: Can the $\gamma_5$ anomaly appear
in the weak interaction?
\par
For this purpose we, at first, use the functional-integral measure method
for the vector theory, considered by K. Fujikawa [6],
and then use this method for the weak
interaction.
\par
At the $\gamma_5$ transformations
$$
\Psi(x)' \rightarrow exp(i \alpha(x) \gamma_5) \Psi
$$
$$
\bar \Psi(x)' \rightarrow exp(i \alpha(x) \gamma_5) \bar \Psi  ,
\eqno(12)
$$
we get the following increment to lagrangian ${\cal L}$
$$
{\cal L} \rightarrow {\cal L} - \partial_\mu \alpha(x)
\bar \Psi \gamma^\mu \gamma_5 \Psi - 2mi \alpha(x) \bar \Psi \gamma_5 \Psi ,
\eqno(13)
$$
where
$$
{\cal L} = \bar \Psi (i\hat D - m) \Psi + (\frac{g^2}{2}) tr F^{\mu \nu}
F_{\mu \nu}
$$
and $\alpha(x)$ is infinitesimal parameter.
\par
Functional- integral measure is defined by the following equation:
$$
d\mu \equiv \prod_x DA_\mu (x) D \bar \Psi(x) D \Psi(x)
\eqno(14)
$$
Under infinitesimal transformations this measure does not remain invariant
and we get (see Appendix and work [6])
$$
d \mu' \rightarrow d \mu exp[i \int \alpha(x) (\frac{1}{8 \pi^2})
tr ^{*}F^{\mu \nu} F_{\mu \nu} dx]  ,
\eqno(15)
$$
where $^{*}F^{\mu \nu} = \epsilon^{\sigma \rho \mu \nu} F_{\sigma \rho}$.
\par
According to the requirement of the measure invariance at this infinitesimal
transformation, we obtain
$$
\partial_\mu (\bar \Psi \gamma^\mu \gamma_5 \Psi) = 2m i \bar \Psi \gamma_5
\Psi
-   i \frac{1}{8 \pi^2} \epsilon^{\sigma \rho \mu \nu} F_{\sigma \rho}
F_{\mu \nu}
\eqno(16)
$$
The second term of the right part of (16) is the $\gamma_5$ anomaly term.
\par
In the case of the weak interaction $\Psi_R = \bar \Psi_R \equiv 0$ and
$$
\bar \Psi \rightarrow \Psi_L  , \Psi \rightarrow \Psi_L  ,
\eqno(17)
$$
then the functional-integral measure is zero (see Appendix)
$$
d \mu \equiv 0  ,
\eqno(18)
$$
and the $\gamma_5$ anomaly term of the right part of Eq. (16) is also zero.
\par
So, we see that in the weak interaction the $\gamma_5$ anomaly does not
appear and the equation type of Eq.(16) for the weak interaction has the
following form:
$$
\partial_\mu (\bar \Psi_L \gamma^\mu \gamma_5 \Psi_L) \equiv 0 .
\eqno(19)
$$

\par
{\bf 3. Conclusion}

\par
In this work it was shown that,
in contrast to the strong and electromagnetic theories,
additive conserved numbers (such as lepton, aromatic and another numbers)
 and $\gamma_5$ anomaly do not appear in the standard weak interaction theory.
It means that in this interaction the additive numbers cannot be conserved.
These results are the consequence of
specific character of the weak interaction: the right components of spinors
 do not participate in this interaction.
The schemes of violation of the aromatic and lepton numbers were considered.\\

\par
{\bf Appendix}
\par
Under the chiral transformation (12)
$$
\Psi(x)' \rightarrow exp(i \alpha(x) \gamma_5) \Psi
$$
the coefficient $a_n$ of the following expansions:
$$
\begin{array}{c}
\Psi(x) = \sum_{n} a_n \phi_n \\
\bar \Psi(x) = \sum_{n} \phi^{+}_n \bar b_n
\end{array}
\eqno(a.1)
$$
$$
d \mu = \prod_{x} [D A_{\mu} (x)] \prod_{m,n} d\bar b_m da_n  ,
$$
(where $\hat D\phi(x) = \lambda_n \phi_n,\qquad \int \phi^{+}_n(x)
\phi_m (x) d^4x = \delta_{n,m}$ and $a_n, b^{+}_m$ are the elements of the
Grassman algebra)\\
are transformed as
$$
{a'}_n = \sum_{m} \int \phi^{+}_n exp(i \alpha(x) \gamma_5) \phi_m dx a_m
= \sum_{m} c_{nm} a_m  .
\eqno(a.2)
$$
Then
$$
\prod_{n} d{a'}_n = (det C_{k,l})^{-1} \prod_{n} da_n
\eqno(a.3)
$$
where
$$
(det C_{k,l})^{-1} = det( \delta_{k,l} + i\int \alpha (x)
\phi^{+}_k(x)\gamma_5 \phi_l(x) dx)^{-1} =
$$
$$
= exp(-i \int \alpha(x) \sum_{k} \phi^{+}_k(x) \gamma_5 \phi_k(x) dx) .
$$
\par
The summation in the exponent of (12) is a bad-defined quantity and
evaluating it by introduction a cutoff $M$ ($\mid \lambda_k \mid \le
M$) we have
$$
\begin{array}{c} \sum_{k} \phi^{+}_n (x) \gamma_5
\phi_k(x) = \lim_{M \to \infty} \sum_{k} \phi^{+}_k \gamma_5
exp\left[-\left({\lambda_k\over M}\right)^2\right] \phi_k(x) = \\
= \lim_{M\to \infty, y \to x} tr \gamma_5
exp\left[-\left({\hat D\over M }\right)^2 \right] \delta(x - y)= \\
= \lim_{M \to \infty, y \to x} \int {dx\over (2\pi)^4} tr \gamma_5
exp\left[- (D^\mu D_\mu + {1\over 4}[\gamma^\mu, \gamma^\nu]F_{\mu\nu})
/ M^2 \right] e^{ik(x - y)} = \\
= \lim_{M \to \infty} {1\over 16} tr \gamma_5  ([\gamma^\mu, \gamma^\nu ]
F_{\mu\nu})^2 {1\over 2M^4} \int {dk\over
{2\pi}^4} exp^{(-{k^2\over M^2})} .
\end{array}
\eqno(a.4)
$$ After the integration one abtains
$$
\sum_{k} \phi^{+}_k(x) \gamma_5 \phi_k(x) =
-{1\over 16 \pi^2} tr ^{*}F^{\mu\nu} F_{\mu\nu} \eqno
(a.5)
$$
The same result one obtain at $b^{+}_n$ transformation, and as a result
one gets eq. (15), i.e.
$$
d\mu' = d\mu e^{\left(i \int \alpha(x) ({1\over 8 \pi^2}) ^{*}F^{\mu\nu}
F_{\mu\nu}\right)} .
$$
It is clear that in the case of the weak interactions, since
$\phi^{+}_{Rk}(x) = \phi_{Rk} \equiv 0$ we have
$$
\sum_{k} \phi^{+}_{Lk}(x) \gamma_5 \phi_{Lk} \equiv 0
\eqno(a.6)
$$

\par
{\bf  References}\\
\par
\noindent
[1] N.N. Bogolubov, D.V. Shirkov, Introd. to the Quantum Field
\par
Theory, M., Nauka, 1986;
\par
G. Kane Modern Elementary Particle Physics, Add. W. P.C., 1987.
\par
\noindent
[2] Kh.M. Beshtoev, JINR Commun. 2-93-44, Dubna, 1993;
\par
JINR Commun., E2-93-167, Dubna, 1993.;
\par
Chinese Journal of Phys. 34 (1996) 979.
\par
\noindent
[3] Kh.M. Beshtoev Kh.M, JINR Commun. E2-94-221, Dubna, 1994.
\par
\noindent
[4] N. Cabibbo, Phys Rev. Lett., 10 (1963) 531.
\par
M. Kobayashi and K. Maskawa, Prog. Theor. Phys., 49 (1973)
\par
652.
\par
\noindent
[5] Kh.M. Beshtoev, JINR, E2-94-293, Dubna, 1994;
\par
   Turkish Journ. of Physics 20 (1996) 1245;
\par
   JINR Commun., E2-95-535, Dubna, 1995;
\par
   JINR Commun., P2-96-450, Dubna, 1996.
\par
   JINR Commun. E2-97-210, Dubna, 1997.
\par
\noindent
[6] K. Fujikawa, Phys. Rev. Lett., 42 (1979) 1195.

\end{document}